\documentclass{article}
\usepackage{algorithm}
\usepackage{algpseudocode}
\usepackage{amsmath}
\usepackage[utf8]{inputenc}
\usepackage{listings}
\usepackage{xcolor}
\usepackage{graphicx}
\usepackage{wrapfig}
\usepackage{fullpage}
\usepackage{changepage}

\usepackage{subcaption}

\usepackage[colorlinks=true,linkcolor=blue,citecolor=blue,urlcolor=blue]{hyperref}
\usepackage{natbib}

\usepackage{color} 

\definecolor{codegreen}{rgb}{0,0.6,0}
\definecolor{codegray}{rgb}{0.5,0.5,0.5}
\definecolor{codepurple}{rgb}{0.58,0,0.82}
\definecolor{backcolour}{rgb}{0.95,0.95,0.92}

\lstdefinestyle{mystyle}{
    backgroundcolor=\color{backcolour},   
    commentstyle=\color{codegreen},
    keywordstyle=\color{magenta},
    numberstyle=\tiny\color{codegray},
    stringstyle=\color{codepurple},
    basicstyle=\ttfamily\footnotesize,
    breakatwhitespace=false,         
    breaklines=true,                 
    captionpos=b,                    
    keepspaces=true,                 
    numbers=left,                    
    numbersep=5pt,                  
    showspaces=false,                
    showstringspaces=false,
    showtabs=false,                  
    tabsize=2
}

\usepackage{listings}
\usepackage{xcolor}

\definecolor{codegreen}{rgb}{0,0.6,0}
\definecolor{codegray}{rgb}{0.5,0.5,0.5}
\definecolor{codepurple}{rgb}{0.58,0,0.82}
\definecolor{backcolour}{rgb}{0.95,0.95,0.95}

\lstdefinestyle{pythonstyle}{
    backgroundcolor=\color{backcolour},   
    commentstyle=\color{codegreen},
    keywordstyle=\color{blue},
    numberstyle=\tiny\color{codegray},
    stringstyle=\color{codepurple},
    basicstyle=\ttfamily\footnotesize,
    breakatwhitespace=false,         
    breaklines=true,                 
    captionpos=b,                    
    keepspaces=true,                 
    numbers=left,                    
    numbersep=5pt,                  
    showspaces=false,                
    showstringspaces=false,
    showtabs=false,                  
    tabsize=2,
    frame=single,
    language=Python
}

\title{A Note on Implementation Errors in Recent\\Adaptive Attacks Against Multi-Resolution Self-Ensembles}
\author{Stanislav Fort\\Google DeepMind}
\date{\today}

\title{\vspace{-1em}A Note on Implementation Errors in Recent\\Adaptive Attacks Against Multi-Resolution Self-Ensembles}
\author{Stanislav Fort \\ \small{\emph{Independent Researcher}}}
\date{January 22, 2025}

\begin{document}

\maketitle

\begin{abstract}
This note documents an implementation issue in recent adaptive attacks (\cite{zhang2024gradientmaskingallatonceensemblev1}) against the multi-resolution self-ensemble defense (\cite{fort2024ensembleeverywheremultiscaleaggregation}). The implementation allowed adversarial perturbations to exceed the standard $L_\infty = 8/255$ bound by up to a factor of 20$\times$, reaching magnitudes of up to $L_\infty = 160/255$. When attacks are properly constrained within the intended bounds, the defense maintains non-trivial robustness. Beyond highlighting the importance of careful validation in adversarial machine learning research, our analysis reveals an intriguing finding: properly bounded adaptive attacks against strong multi-resolution self-ensembles often align with human perception, suggesting the need to reconsider how we measure adversarial robustness.

\end{abstract}

\begin{figure}[htbp]
    \centering
    \begin{subfigure}[b]{0.48\textwidth}
        \centering
        \includegraphics[width=\textwidth]{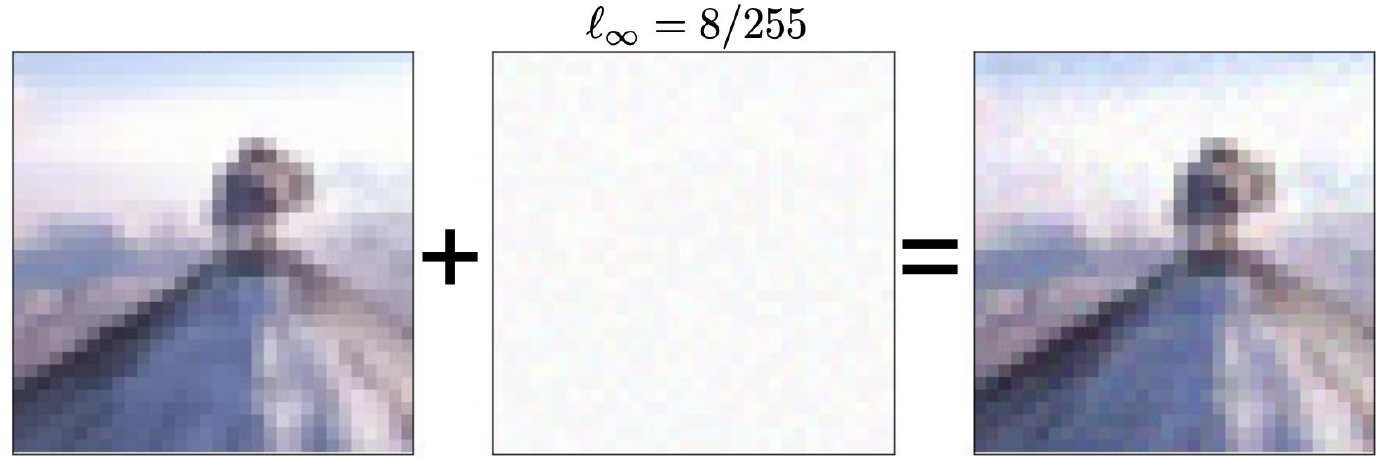}
        \caption{A standard adversarial attack with $L_\infty = 8/255$ perturbation. The perturbation is constrained within acceptable bounds, resulting in visually subtle changes.}
        \label{fig:attack8}
    \end{subfigure}
    \hfill
    \begin{subfigure}[b]{0.48\textwidth}
        \centering
        \includegraphics[width=\textwidth]{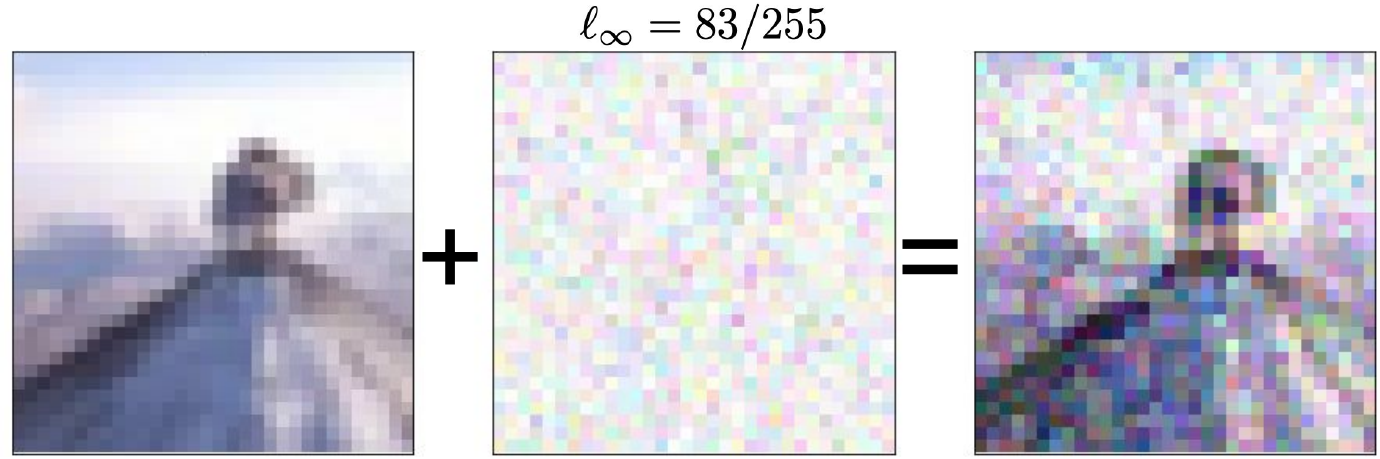}
        \caption{An attack from \citet{zhang2024gradientmaskingallatonceensemblev1} reaching $L_\infty = 83/255$ due to an implementation error. The perturbation is clearly excessive}
        \label{fig:attack83}
    \end{subfigure}
    \caption{Comparison of standard adversarial perturbations and implementation-error perturbations. In \citet{zhang2024gradientmaskingallatonceensemblev1}, the authors used an adaptive adversarial attack technique that, due to an implementation error and unbeknownst to them, exceeded the standard attack bound of $L_\infty = 8/255$ by a factor of over 10$\times$. The left panel shows a standard attack within accepted bounds, while the right panel showcases a real adversarial perturbation generated by the authors' code that significantly exceeds these bounds. The excessive perturbation becomes clearly visible to human observers, indicating a significant departure from the standard evaluation protocol where perturbations should be imperceptible.}
    \label{fig:attacks-comparison}
\end{figure}

\vspace{-0.5cm}
\section{Introduction}
Recent work by \citet{zhang2024gradientmaskingallatonceensemblev1} reported breaking the multi-resolution self-ensemble defense (introduced in \citet{fort2024ensembleeverywheremultiscaleaggregation}) through an adaptive attack strategy. Their approach used multiple restarts to gradually refine adversarial perturbations, reporting a reduction in the defense’s adversarial accuracy to near zero. However, our subsequent investigation revealed that these results stemmed from perturbations, unbeknownst to the authors, significantly exceeding the standard $L_\infty = 8/255$ bound, widely used in the field. See Figure~\ref{fig:attacks-comparison} for an example of an $L_\infty \approx 83/255$ attack generated by the authors’ adaptive attack code.

This implementation issue, which allowed perturbations to reach magnitudes of up to $L_\infty = 160/255$, fundamentally alters the interpretation of the reported results. When properly constrained within the standard bounds, the defense maintains meaningful robustness. Beyond documenting this specific case, our technical note raises broader questions about validation practices in adversarial machine learning research and how we measure adversarial robustness.

Moreover, our investigation uncovered an unexpected finding: when properly bounded adaptive attacks do succeed against strong multi-resolution self-ensembles, they often produce perturbations that align with human perception. This observation challenges the conventional assumption that $L_\infty = 8/255$ perturbations should never affect human classification, suggesting the need for more nuanced evaluation metrics in adversarial robustness research. We demonstrate this in Figure~\ref{fig:interpretable-attacks} on two examples.

\section{Documentation of the Error}

\subsection{Discovery Timeline}
The implementation error was identified during our validation efforts before the Neural Information Processing Systems (NeurIPS) 2024 conference on December 10, 2024. Initial attempts to reproduce the reported attacks using a custom implementation with rigorous perturbation tracking led to inconsistent results. After examining the authors' private GitHub repository\footnote{\url{https://github.com/ethz-spylab/attack_ens_everything}}, we identified that perturbations were accumulating beyond the standard $L_\infty = 8/255$ bound.
The original authors promptly acknowledged this issue. The first author confirmed our findings via email:

\begin{quote}
``[...] Thank you for bringing that to my attention. After reviewing the code, I agree with your assessment. [...]''\\
-- Jie Zhang, December 10, 2024, private email correspondence
\end{quote}
The senior author subsequently acknowledged this publicly:
\begin{quote}
``This was an unfortunate mistake, sorry about that. [...]''\\
-- Florian Tramèr, December 12, 2024, \citet{tramer2024tweet}
\end{quote}

\subsection{Technical Analysis}

The implementation error stems from a subtle but consequential issue in how perturbations compound across iterations. While each individual perturbation was correctly bounded by $L_\infty = 8/255$, the implementation allowed these perturbations to accumulate across iterations rather than maintaining the bound relative to the original image.
The effect can be understood through a simple progression:
\begin{enumerate}
    \item First iteration: Applies a perturbation bounded by $L_\infty \le 8/255$
    \item Second iteration: Takes the perturbed image as the new baseline and adds another  $L_\infty \le 8/255$ perturbation
    \item This process continues for up to 20 iterations
\end{enumerate}

Our replication of the code confirmed this accumulation pattern: we found that in the first iteration, the $L_\infty \le 8/255$ as expected. However, at iteration 2, we saw $L_\infty \approx 15.8/255$, iteration 3 reached $L_\infty \approx 23.6/255$ and iteration 4 was around $L_\infty \approx 30.54/255$, following an approximate $L_\infty(n) = 8n / 255$ pattern, as expected.

The code-level analysis reveals the source of this behavior. While the core PGD (Projected Gradient Descent) implementation correctly enforces the perturbation bound $\varepsilon$ within each iteration, the multi-round implementation incorrectly uses the perturbed output of the previous round as the starting point instead of the original, unperturbed image.

The relevant code sections highlight this issue:

\begin{lstlisting}[caption={PGD attack implementation}]
# Inside _pgd_attack function - correct bounds within single PGD
eta = step_size * X_pgd.grad.data.sign()
X_pgd = Variable(X_pgd.data + eta, requires_grad=True)
eta = torch.clamp(X_pgd.data - X.data, -epsilon, epsilon)
X_pgd = Variable(X.data + eta, requires_grad=True)
\end{lstlisting}

This correctly enforces the $L_\infty$ bound of $\epsilon$ within each PGD iteration by: 1) Computing the signed gradient step, 2) Applying the step, 3) Clamping the total perturbation relative to the original input \texttt{X}, and 4) Reconstructing the perturbed input from the original plus the clamped perturbation.

However, the multi-round adaptive attack implementation introduces a critical flaw:

\begin{lstlisting}[caption={Adaptive attack implementation with accumulation issue}]
# In adaptive_attack function - problematic accumulation across rounds
for round_num in range(args.num_rounds):
    # ... identify failed samples ...
    for i in range(0, len(saved_adv_images_failed), args.bs):
        X_batch = saved_adv_images_failed[i:i + args.bs]
        _, _, X_pgd, _ = _pgd_attack(model_target, X_batch, y_batch, 
                                    num_steps=pgd_steps, num_eot=10)
        saved_adv_images_np[failed_indices[i:i + args.bs]] = \
            X_pgd.cpu().detach().numpy()
\end{lstlisting}

This implementation meant that after $n$ rounds, the actual $L_\infty$ norm of the perturbation could be up to $n$ times larger than the intended bound $\varepsilon$. With the authors' default setting of $n = 20$ rounds, this led to perturbations that simply overwhelmed the original image content with noise.

\section{Properly Bounded Adaptive Attacks and Human Perception}
\begin{figure}[htbp]
\vspace{-0.5cm}
    \centering
    \begin{subfigure}[b]{0.48\textwidth}
        \centering
        \includegraphics[width=\textwidth]{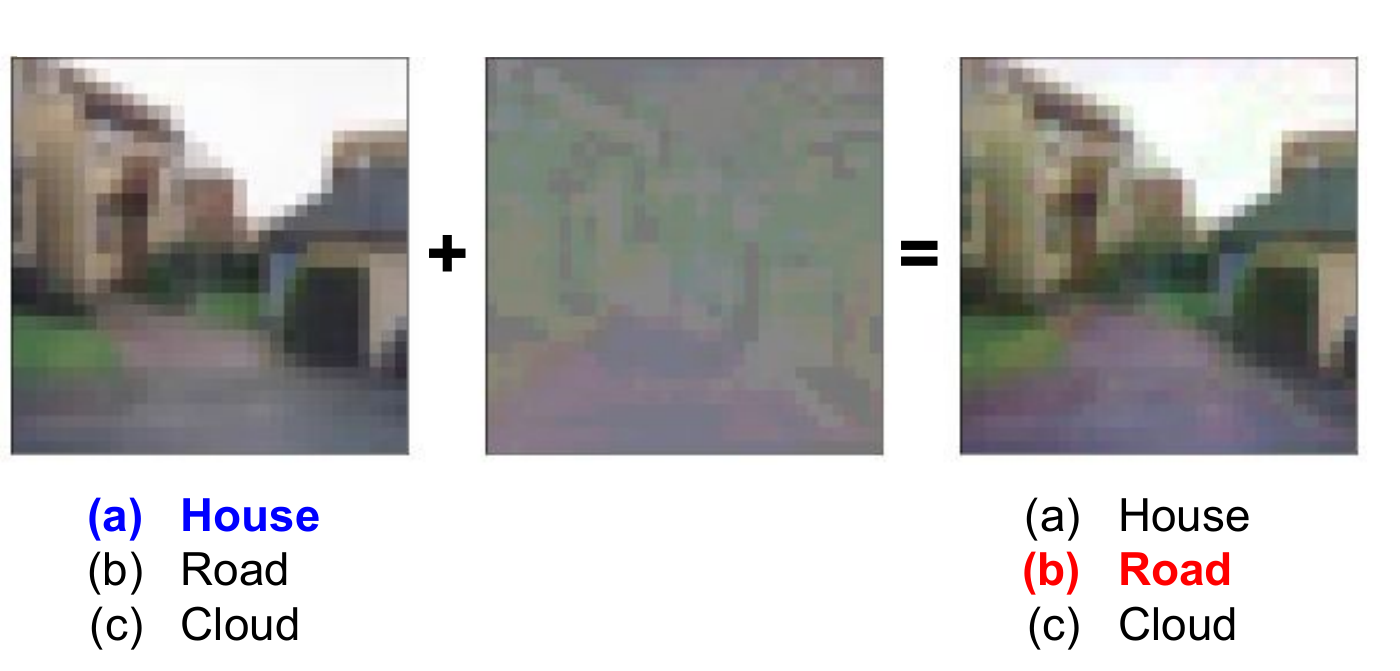}
        \caption{An adaptive attack changing \textit{house} to \textit{road}}
        \label{fig:house-to-road}
    \end{subfigure}
    \hfill
    \begin{subfigure}[b]{0.48\textwidth}
        \centering
        \includegraphics[width=\textwidth]{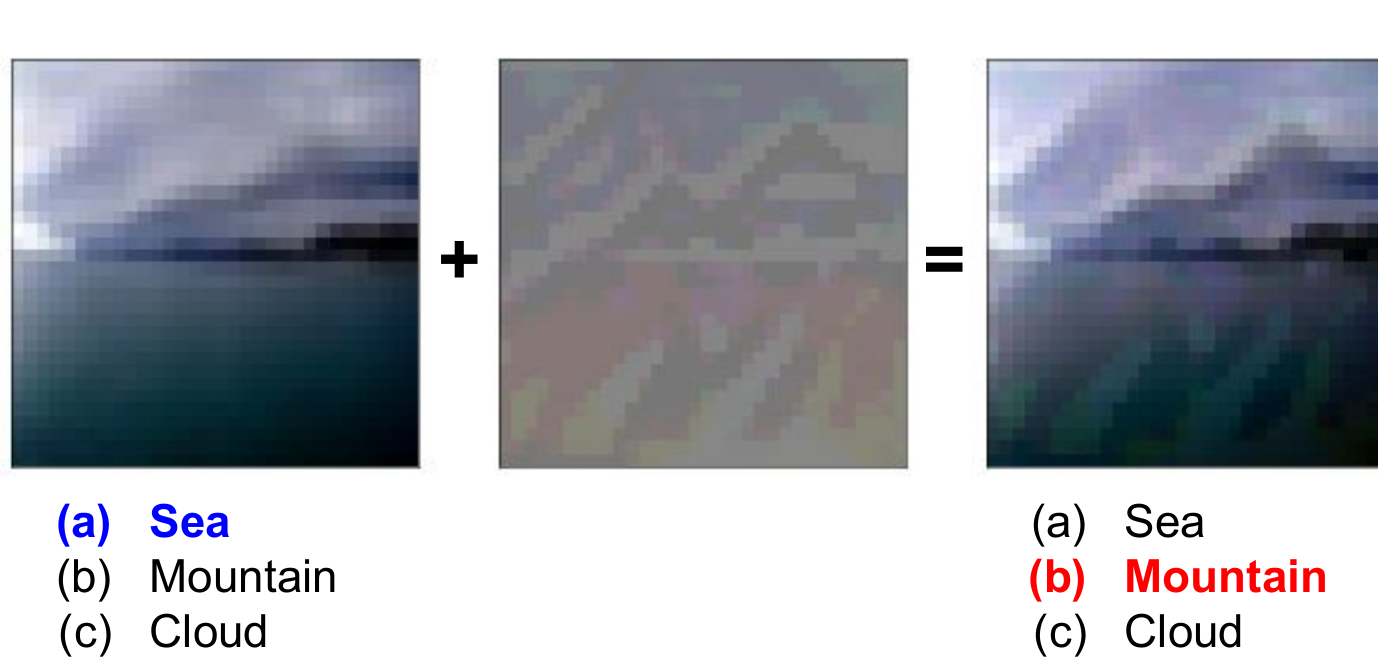}
        \caption{An adaptive attack changing \textit{sea} to \textit{mountain}}
        \label{fig:sea-to-mountain}
    \end{subfigure}
    \caption{Standard adversarial accuracy benchmarks assume that humans will not perceive a different class under $L_\infty = 8/255$ perturbations, but our properly bounded adaptive attacks challenge this assumption. Here we show two "successful" adaptive attacks on a multi-resolution self-ensemble, changing the model's decision from house to road on the left, and sea to mountain on the right. When participants of a lecture at EPFL \citep{fort2024epfllecture} presented these attacked images and a limited set of classes to choose from, they agreed with the target label, being "confused" in the same way the model was. This shows a remarkable alignment of the model with human perception. Given that this is traditionally labeled as a "successful" attack, we believe that we need to reconsider how we establish adversarial robustness.}
    \label{fig:interpretable-attacks}
\end{figure}

We implemented adaptive attacks against the multi-resolution self-ensemble while maintaining the standard $L_\infty = 8/255$ bound. Our implementation used a multi-resolution self-ensemble based on all 54 layers of the \verb|ResNet152| architecture described in \citet{fort2024ensembleeverywheremultiscaleaggregation}, a significantly stronger model than the demonstration version used in \citet{zhang2024gradientmaskingallatonceensemblev1} based on a quick demonstration public Colab notebook\footnote{\url{https://github.com/stanislavfort/ensemble-everything-everywhere}}. Under these conditions, we measured a preliminary adversarial accuracy of $>20\%$ on CIFAR-100 under $L_\infty$ attacks, in contrast to the near-zero robustness reported in the original work that relied on substantially larger perturbations.

However, our investigation revealed an unexpected and intriguing phenomenon. When examining cases where our properly bounded adaptive attacks successfully fooled the multi-resolution self-ensemble, we found that these perturbations not infrequently affected human perception in similar ways. This observation challenges a fundamental assumption in adversarial robustness research: that $L_\infty = 8/255$ perturbations should never change the perceived class of an image.

We demonstrated this effect during a lecture at the EPFL AI Center in November 2024 \citep{fort2024epfllecture}. When presented with two "successfully" attacked images (Figure~\ref{fig:interpretable-attacks}) and a limited set of relevant classes to choose from, the audience consistently misclassified the images in alignment with the model's predictions. For example, the perturbed \textit{house} image was overwhelmingly classified as \textit{road} by both the model and human observers, while the \textit{sea} image was consistently perceived as \textit{mountain}.

This alignment between model and human perception raises important questions about how we evaluate adversarial robustness. Current benchmarks label such cases as "successful attacks" based on the assumption that $L_\infty = 8/255$ perturbations should be imperceptible. However, our findings suggest that some of these "successful" attacks may actually represent cases where the perturbation has genuinely altered the perceptual features that both humans and models use for classification.

These observations suggest that we may need to refine our approach to measuring adversarial robustness, particularly as models become more sophisticated and better aligned with human visual processing. Instead of treating all prediction changes under $L_\infty = 8/255$ as failures of robustness, we might need to consider whether the perturbation has meaningfully altered the perceptually relevant features of the image.

\vspace{-0.2cm}
\section{Conclusion}

While the multi-resolution self-ensemble defense shows some vulnerability to properly-bounded adaptive attacks, its robustness remains non-trivial and significantly higher than reported in \citet{zhang2024gradientmaskingallatonceensemblev1}. Their originally reported near-zero robustness result relied on perturbations that exceeded the standard $L_\infty = 8/255$ bound by up to a factor of 20$\times$ (i.e. up to $L_\infty = 160/255$), highlighting the importance of careful implementation validation in adversarial machine learning research.

More intriguingly, our analysis revealed that when using a strong model configuration, the defense often fails in ways that align precisely with human perception. When properly bounded adaptive attacks successfully fool the multi-resolution self-ensemble, they frequently produce perturbations that induce similar "misclassifications" in human observers. This finding challenges our current approach to measuring adversarial robustness, which assumes that $L_\infty = 8/255$ perturbations should never affect human classification decisions.

This experience reinforces the critical role of implementation validation in adversarial machine learning research while also suggesting that we reconsider how we evaluate and define adversarial robustness.

\bibliographystyle{unsrtnat}
\vspace{-0.3cm}
\bibliography{main}

\end{document}